\documentclass[conference]{IEEEtran}
\IEEEoverridecommandlockouts
\usepackage{cite}
\usepackage{amsmath,amssymb,amsfonts}
\usepackage{algorithmic}
\usepackage{graphicx}
\usepackage{textcomp}
\usepackage{xcolor}
\usepackage[a4paper, total={184mm,239mm}]{geometry}
\def\BibTeX{{\rm B\kern-.05em{\sc i\kern-.025em b}\kern-.08em
    T\kern-.1667em\lower.7ex\hbox{E}\kern-.125emX}}
\usepackage{booktabs}   
\usepackage{bm}
\usepackage{subfigure}
\usepackage{algorithm}
\usepackage{multirow}
\usepackage{upgreek}
\usepackage{siunitx} 
\usepackage{array} 
\usepackage[T1]{fontenc}
\usepackage{courier}
\usepackage{float}
\usepackage{threeparttable}
\usepackage{url}
\usepackage{hyperref}
\newcommand{\BR}[1]{\textcolor{brown}{#1}}

\def\@IEEEauthorblockNstyle{\small\normalfont\sublargesize}
\def\@IEEEauthorblockAstyle{\small\normalfont\itshape\normalsize}

\def\@IEEEauthorblockNtopspace{0.5ex}
\def\@IEEEauthorblockAtopspace{0.5ex}

\def\@IEEEauthorblockNinterlinespace{2.0ex}
\def\@IEEEauthorblockAinterlinespace{2.0ex}

\def\@IEEEauthorblockconfadjspace{-10em}

\begin{document}

\title{Timing-Driven Global Placement \\by Efficient Critical Path Extraction}

\author{
\IEEEauthorblockN{Yunqi Shi$^{1,2,3}$, Siyuan Xu$^{3}$, Shixiong Kai$^{3}$, Xi Lin$^{1,2}$, Ke Xue$^{1,2}$, Mingxuan Yuan$^{3}$, Chao Qian$^{1,2}$}
\IEEEauthorblockA{$^{1}$National Key Laboratory for Novel Software Technology, Nanjing University, China \\
$^{2}$School of Artificial Intelligence, Nanjing University, China \\
$^{3}$Huawei Noah's Ark Lab, China \\
\{shiyq, qianc\}@lamda.nju.edu.cn}
}

\maketitle

\begin{abstract}
Timing optimization during the global placement of integrated circuits has been a significant focus for decades, yet it remains a complex, unresolved issue. Recent analytical methods typically use pin-level timing information to adjust net weights, which is fast and simple but neglects the path-based nature of the timing graph. The existing path-based methods, however, cannot balance the accuracy and efficiency due to the exponential growth of number of critical paths. In this work, we propose a GPU-accelerated timing-driven global placement framework, integrating accurate path-level information into the efficient DREAMPlace infrastructure. It optimizes the fine-grained pin-to-pin attraction objective and is facilitated by efficient critical path extraction. We also design a quadratic distance loss function specifically to align with the RC timing model. Experimental results demonstrate that our method significantly outperforms the current leading timing-driven placers, achieving an average improvement of 40.5\% in total negative slack (TNS) and 8.3\% in worst negative slack (WNS), as well as an improvement in half-perimeter wirelength (HPWL).
\end{abstract}


\section{Introduction}

In the field of very-large-scale integration (VLSI) design, the placement process is critical as it forms the bridge between logical design and physical layout~\cite{chu2009placement,wang2024benchmarking}. Traditional placement methods, while focusing on minimizing wirelength and reducing routing congestion, only implicitly address timing metrics~\cite{caldwell1998wirelength}, which may fail to satisfy the strict timing requirements of modern, large-scale chip designs. Direct optimization of timing is essential but typically demands considerable computational resources and turn-around time, emphasizing the need for more efficient timing-driven placement methods to improve design cycles and ensure timing closure.

Modern placement algorithms often consist of three main stages: global placement, legalization, and detailed placement~\cite{markov2012progress}. Global placement distributes cells across the target layout, balancing the wirelength and density. The coarse result is then refined by legalization and fine-tuned by detailed placement. Among these three stages, global placement plays a crucial role in determining the overall distribution of cells, significantly influencing the quality of the final placement, including timing. As a result, timing-driven placement (TDP) for global placement has been extensively studied, focusing on optimizing key timing metrics such as total negative slack (TNS) and worst negative slack (WNS).

Such TDP techniques basically have three components: foundational placement algorithms, timing analysis, and interfaces between them~\cite{pan2008timing}. The first component utilizes traditional global placement engines, which primarily focus on optimizing the trade-offs between wirelength and density. The second component involves either internal or external timing engines that assess the current layout of the placement to provide essential timing data, such as critical path delays or pin slacks. The third component translates timing metrics into certain weights or constraints to drive the foundational placement engines. Depending on the method of handling timing information, TDP techniques can be broadly categorized into two types: net-based and path-based approaches. 

\textbf{Net-based} methods use timing analysis to adjust net weights~\cite{burstein1985timing,Dunlop1984chip,chang2002net,eisenmann1998generic,obermeier2004quadratic} or net constraints~\cite{kahng2011vlsi,luk1991fast,gao1992performance} either dynamically or statically, indirectly guiding the placement to focus on critical nets. Since traditional placement algorithms primarily focus on minimizing wirelength, which inherently involves net considerations, only minimal modifications are required to adapt these for a timing-driven approach. Recently, Liao et al. upgraded the advanced nonlinear placer, DREAMPlace~\cite{lin2019dreamplace}, to its timing-driven version 4.0~\cite{liao2023dreamplace}. This new version dynamically adjusts net weights, utilizing a momentum-guided mechanism that interacts with a timing analysis engine, enhancing its focus on timing optimization. 

\textbf{Path-based} methods~\cite{chowdhary2005accurately,jackson1989performance,swartz1995timing} directly address paths extracted from the timing graph, typically formulated as a mathematical programming problem. These methods maintain an accurate view of timing during the optimization~\cite{chang1994practical}, thereby often ensuring high-quality results. However, they frequently encounter scalability issues as the number of paths grows exponentially with the increase of design size~\cite{liao2023dreamplace}. Recently, Guo and Lin introduced a novel differentiable-timing-driven placement framework~\cite{guo2022differentiable}, which incorporates a GPU-accelerated, differentiable timing engine into DREAMPlace, enabling efficient path-based analysis. This approach not only achieves state-of-the-art performance but also operates at competitive speeds, addressing traditional scalability challenges effectively.

Despite significant advancements, the timing-driven placement problem remains largely unsolved. Net-based approaches often suffer from an indirect optimization objective and underutilized timing information. For path-based methods, although Guo and Lin~\cite{guo2022differentiable} have somewhat addressed the scalability issue, their approach potentially compromises accuracy by smoothing timing metrics. In this work, we introduce a timing-driven global placement framework that incorporates a fine-grained pin-to-pin attraction quadratic distance loss, directly targeting timing metrics. This is complemented by a path-level timing analysis module that extracts critical paths efficiently. We outline the key contributions as follows:
\begin{itemize}
    \item We develop a GPU-accelerated, timing-driven placement flow that optimizes pin-to-pin attraction on critical paths, based on the leading placer DREAMPlace 4.0~\cite{liao2023dreamplace}. Our code is available at \url{https://github.com/lamda-bbo/Efficient-TDP}.
    \item We introduce an efficient critical path extraction method that captures comprehensive timing information, enabling timing optimization at a high speed—achieving a 6$\times$ speed improvement over the default timer~\cite{huang2020opentimer}.
    \item We design a quadratic Euclidean distance loss for pin-to-pin attraction, which is closely aligned with timing metrics and significantly contributes to the superior performance, showing 50\% (30\%) improvements on TNS (WNS) compared to other distance metrics.
    \item Experimental results on the ICCAD2015 contest benchmark suites~\cite{kim2015iccad} show that we can achieve about 60\% (30\%) improvements on TNS (WNS), compared to DREAMPlace 4.0~\cite{liao2023dreamplace} and about 50\% (10\%) improvements on TNS (WNS), compared to Guo and Lin's work~\cite{guo2022differentiable}.
\end{itemize}

The rest of the paper is organized as follows: Section~\ref{sec:preliminaries} gives the preliminaries for timing-driven global placement. Section~\ref{sec:our_algorithm} presents details of our proposed timing-driven placement flow. Section~\ref{sec:experiments} provides empirical studies and discussions. Section~\ref{sec:conclusion} concludes this paper.

\section{PRELIMINARIES}\label{sec:preliminaries}


\subsection{Nonlinear Global Placement}

Global placement is a critical phase in physical design, aiming to determine the locations of millions of cells within a specified chip layout. The goal of optimization is to minimize the wirelength connecting all relevant components while adhering to density constraints. To facilitate efficient optimization, the constrained problem is transformed into an unconstrained nonlinear optimization problem:

\begin{equation}\label{eq:obj}
    \min_{\bm{x}, \bm{y}} \sum_{e\in E} WL_e(\bm{x}, \bm{y}) + \lambda \cdot D(\bm{x}, \bm{y}), 
\end{equation}
where $\bm{x}$, $\bm{y}$ are the cell locations, $E$ represents the set of all nets, $WL$ is typically a smoothed half-perimeter wirelength (HPWL) function (e.g., weighted-average method~\cite{hsu2011tsv}), $D$ is a density metric, and $\lambda$ is the density penalty factor.

\subsection{Static Timing Analysis}

Static timing analysis (STA)~\cite{circuits1999timing} evaluates circuit timing by modeling it as a directed acyclic graph, where edges represent timing arcs that indicate signal propagation directions. In this graph, a timing path starts from a source and ends at a sink, with the arrival time $Arr$ propagated forward and the required arrival time $Req$ backward along the path. The difference between these times at any point $t$ defines the slack:

\begin{equation}
Slack(t) = Req(t) - Arr(t).
\end{equation}

A negative slack at an endpoint indicates a timing violation, necessitating further optimization. To quantify these violations, the metrics worst negative slack (WNS) and total negative slack (TNS) are used. WNS identifies the largest magnitude of violation, while TNS sums all the negative slacks:

\begin{equation}
\text{WNS} = \min\nolimits_{t \in V} Slack(t),
\end{equation}
\begin{equation}
\text{TNS} = \sum\nolimits_{t \in V} Slack(t),
\end{equation}
where $V$ represents the set of all violated endpoints. If $V = \emptyset$, both WNS and TNS are zero, indicating that all timing constraints are met.

\subsection{Timing-Driven Placement by Net Weighting} \label{subsec:TDP_by_net_weighting}
With WNS and TNS defined to measure the timing performance, vanilla nonlinear placement may be inefficient in optimizing these metrics, because wirelength does not directly target timing metrics. An intuitive idea is dynamically adjusting weight of nets using timing information. Among timing-driven placers, DREAMPlace 4.0~\cite{liao2023dreamplace} is the state-of-the-art open-source implementation. Built upon DREAMPlace~\cite{lin2019dreamplace}, version 4.0 makes the most of the GPU acceleration framework and integrates the popular open-source timer OpenTimer~\cite{huang2020opentimer} for STA, further extracting timing information for net weighting. The nonlinear placement objective (\ref{eq:obj}) is thus formulated as:

\begin{equation}
    \min_{\bm{x}, \bm{y}} \sum_{e\in E} w_e \cdot WL_e(\bm{x}, \bm{y}) + \lambda \cdot D(\bm{x}, \bm{y}), 
\end{equation}
where $w_e$ is the weight assigned to net $e$. Nets that are timing critical are assigned larger weights.

\section{Our Algorithm} \label{sec:our_algorithm}

\definecolor{mycolor}{RGB}{255,196,4}
\begin{figure}[t]
\centering
\includegraphics[width=0.46\textwidth]{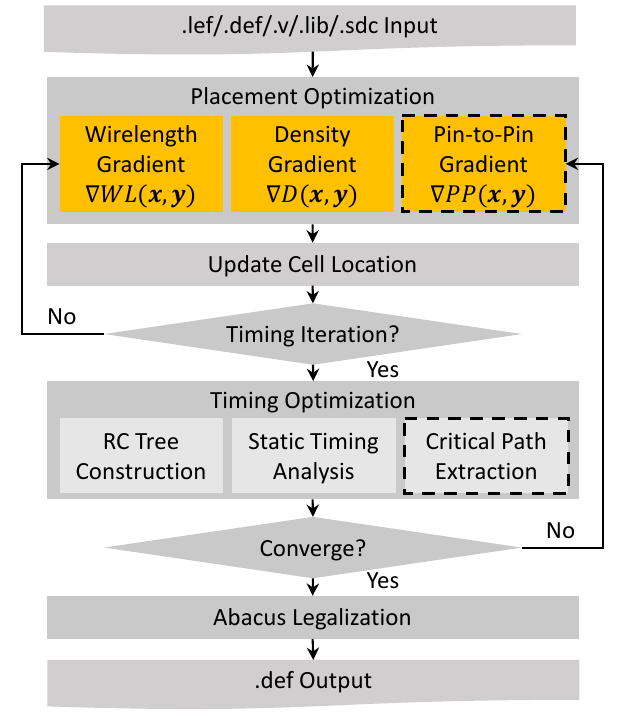}
\caption{Our timing-driven placement flow enabling GPU-acceleration. Gradients in \colorbox{mycolor}{orange} are propagated on GPU.}
\label{fig:flow}
\end{figure}

In this section, we propose a GPU-accelerated timing-driven global placement framework shown in Fig.~\ref{fig:flow}. We introduce a fine-grained pin-to-pin attraction objective, which directly targets timing metrics, facilitated by an efficient critical path extraction method and a quadratic Euclidean distance loss.

\subsection{Fine-Grained Weighting Scheme}

As described in Sec.\ref{subsec:TDP_by_net_weighting}, traditional net weighting methods aim to enhance the timing performance by assigning additional weights to critical nets. However, given the complexity of modern designs, which often feature large fan-out nets and shared data paths, this approach has notable shortcomings. Specifically, it may apply unnecessary weights to non-critical pin pairs and overlook the effects of path-sharing, which disable efficient optimization of timing performance.

To address these issues, we propose incorporating pin-to-pin attraction as a fine-grained objective, replacing the traditional method of applying extra net weights for timing optimization. The revised objective function is presented as follows:
\begin{equation}\label{eq:obj_modi}
    \min_{\bm{x}, \bm{y}} \sum_{e\in E} WL_e(\bm{x}, \bm{y}) + \lambda \cdot D(\bm{x}, \bm{y}) + \beta \cdot PP(\bm{x}, \bm{y}), 
\end{equation}
where $\beta$ represents the penalty multiplier, and $PP$ denotes the pin-to-pin attraction loss. 

\begin{figure}[t]
\centering
\includegraphics[width=0.48\textwidth]{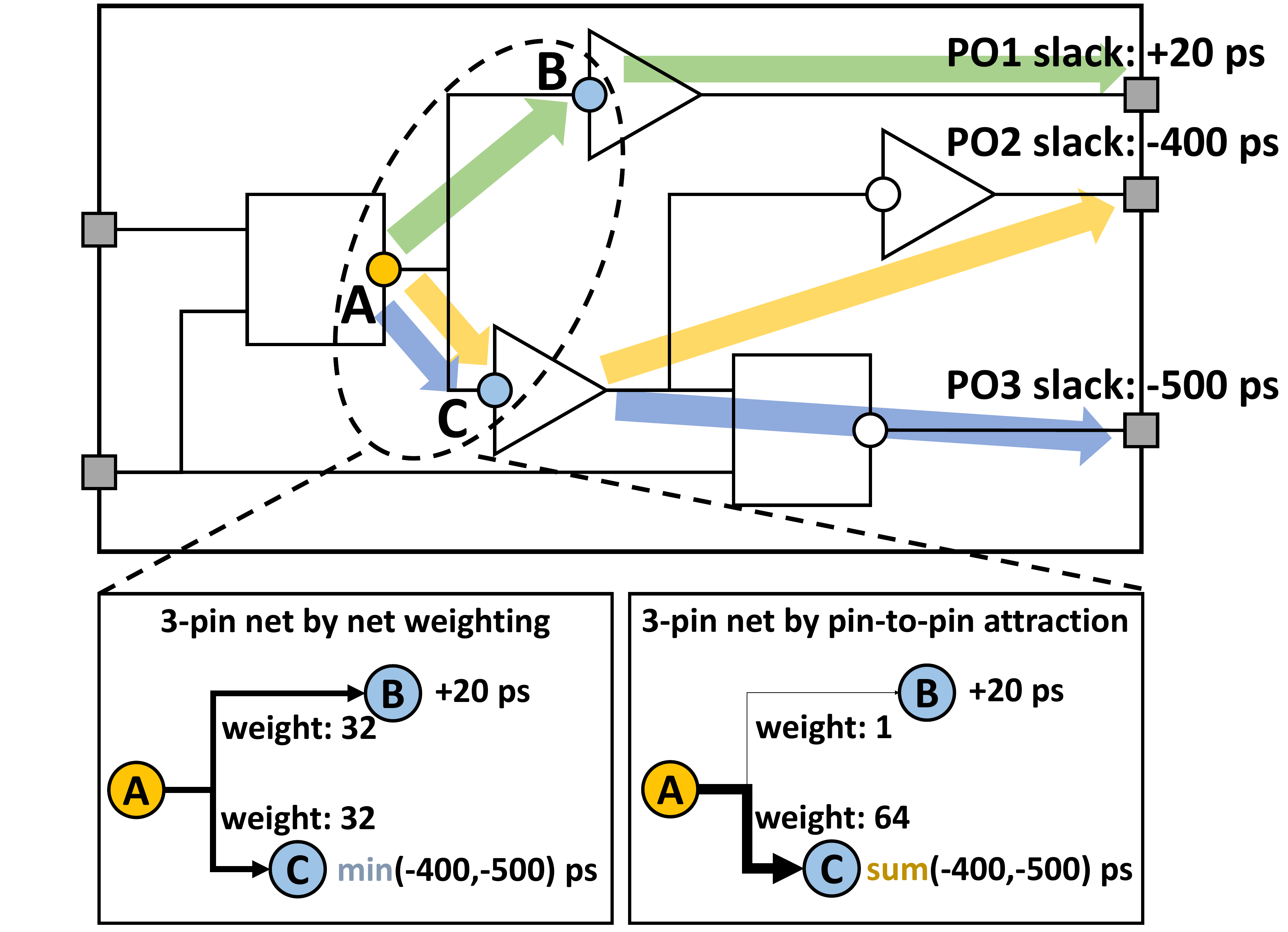}
\caption{Illustration of traditional net weighting and pin-to-pin attraction.}
\label{fig:pin2pin_attraction}
\end{figure}

Pin-to-pin attraction describes an attractive force that brings pins on critical path closer together, thereby reducing wire delay and improving timing performance. Fig.~\ref{fig:pin2pin_attraction} compares the traditional net weighting scheme with the pin-to-pin attraction model for a three-pin net. We illustrate this with an example comprising three timing paths, indicated by green, yellow, and blue arrows. Traditionally, net weighting for timing optimization typically involves assigning a substantial weight to both pins B and C, based on the worst pin slack within the net (i.e., pin C in this case). But the weight is unnecessary for pin B, as positive slacks are disregarded in timing metrics. Furthermore, this heavy weighting could compromise the wirelength of other nets, potentially creating new critical paths. Additionally, pin C's slack is determined by the worst slack of the paths it belongs to, calculated as \texttt{min(-400,-500)}, which ignores the effects of path-sharing due to the nature of pin-level timing analysis. In contrast, the pin-to-pin attraction method assigns weights selectively to critical pin pairs based on their individual slacks, offering a more refined control that benefits both overall timing and wirelength. What's more, we can analyse critical paths one by one, thus define the pin C's slack as \texttt{sum(-400,-500)}, taking path-sharing into consideration.

Although a similar concept of ‘pin-to-pin attraction’ has been previously described as ‘virtual path’ in the literature~\cite{lin2024electrostatics, chen2022virtual}, the advantages of fine-grained weighting and path-sharing cannot be realized without extracting critical timing paths efficiently. This enables proper weights assigned to those critical pin pairs. This vital aspect has been insufficiently explored in prior studies due to its complexity and the potential for an exponential increase in the number of paths as chip scale grows.

\subsection{Critical Path Extraction}

\begin{table*}[ht]

\centering
\begin{threeparttable}
\caption{Timing statistics comparison among various critical path extraction methods.}
\begin{tabular}{@{}lccccc@{}}
\toprule
\textbf{Command}                          & \textbf{Complexity} & \textbf{Number of Paths} & \textbf{Number of Endpoints} & \textbf{Number of Pin Pairs} & \textbf{Time} (\si{sec}) \\ \midrule
\texttt{report\_timing(26300)}            &  $O(n^2)$                   & 26300               & 6                       & 748                           & 41.64              \\
\texttt{report\_timing(263000)}           &  $O(n^2)$                   & 263000              & 20                      & 2538                         & 146.70             \\
\texttt{report\_timing\_endpoint(26300,1)}  &   $O(n\times k)$                  & 26300               & 26300                  & 62811                        & 7.00               \\
\texttt{report\_timing\_endpoint(26300,10)} &   $O(n\times k)$                  & 135705*              & 26300                  & 93740                        & 21.46              \\ \bottomrule
\end{tabular}
\label{table:timing_statistics}
\begin{tablenotes}
\small
\item * The number here is less than 263000, as not all endpoints can extract 10 critical paths.
\end{tablenotes}
\end{threeparttable}
\end{table*}

To address the necessity of efficiently extracting path-level timing information, we integrate OpenTimer~\cite{huang2020opentimer}, a leading timing engine widely adopted by open-source projects and adapted from DREAMPlace 4.0~\cite{liao2023dreamplace}. OpenTimer provides an advanced feature, \texttt{report\_timing(n)}, which identifies the worst $n$ endpoints based on slack and retrieves the $n$ worst critical paths for each, resulting in $n^2$ paths from which the top $n$ worst paths are selected. While \texttt{report\_timing(n)} is effective for identifying critical paths when $n$ is small (e.g.,~1), allowing quick and detailed analysis of specific paths, its efficiency decreases as $n$ increases due to the quadratic growth in analyzed paths. Additionally, the extracted paths tend to concentrate on a few critical endpoints, which does not align well with the TNS metric that requires summing negative slacks across all endpoints. 

To address the afore-mentioned problem, we present the \texttt{report\_timing\_endpoint(n,k)} method for critical path extraction. Here $n$ represents the number of most critical endpoints we want to investigate and $k$ means the number of critical paths we extract for each endpoint. The method returns $n\times k$ paths that ensure each mentioned endpoint is properly covered, thus comprehensively reflecting the timing issue of the entire chip and directly targeting the TNS metric. 

Table~\ref{table:timing_statistics} details the timing analysis for the \texttt{superblue1} case~\cite{kim2015iccad} using different methods. Initially, we identify a total of 26,300 failing endpoints. Employing OpenTimer's \texttt{report\_timing(26300)}, we find that out of these paths, only 6 unique endpoints and 748 unique pin pairs are extracted, which significantly deviates from the TNS metric's requirements that each failing endpoint's slack should be considered. Even increasing the path count to $26300 \times 10$, the extracted unique endpoints and pin pairs still fall short of the necessary criteria for a comprehensive TNS evaluation. In contrast, our method, \texttt{report\_timing\_endpoint(26300,1)}, efficiently covers all endpoints and includes a broader range of pin pairs. Increasing the number of paths per endpoint to 10 triples the time while only increasing the number of pin pairs by 1.5 times, indicating that the former setting is sufficient for effective optimization. Further empirical evaluations and discussions are available in Table~\ref{table:ablation} and Sec.~\ref{sec:ablation}.

\subsection{Quadratic Euclidean Distance Loss}\label{sec:quadratic_loss}

\begin{figure}[t]

\centering
\includegraphics[width=0.48\textwidth]{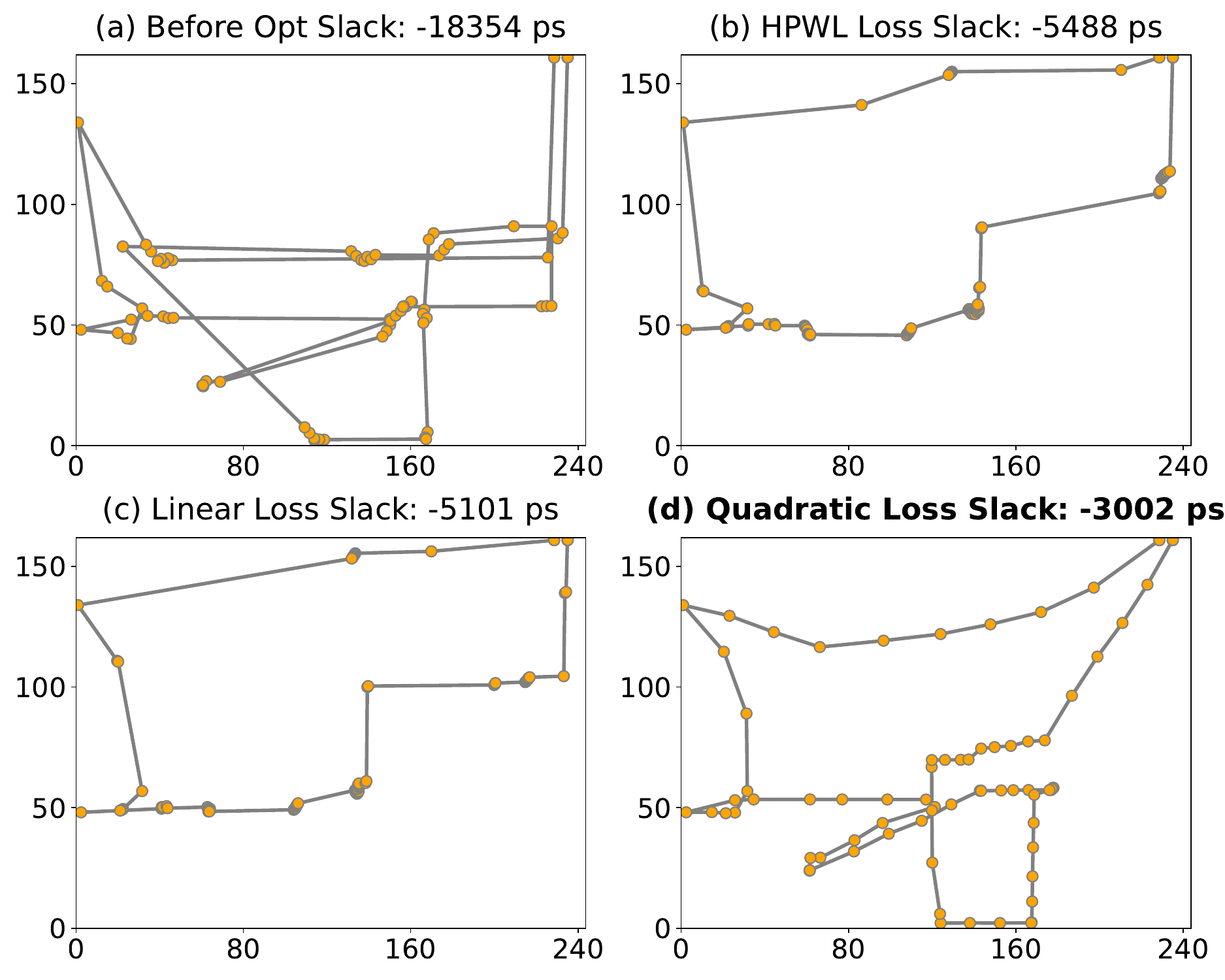}
\caption{Visualization of a specific critical path optimized using different distance losses. The slack of each path is given on the top of each figure.}
\label{fig:critical_path}
\end{figure}

To achieve effective optimization, it is desirable to design a loss function that aligns well with the final timing metrics. In practice, the RC delay model is fairly sufficient and widely adopted. Given the distributed RC network of a net, the delay from the net source $s$ to sink $t$ can be calculated as follows:
\begin{equation}
\text{Delay}_{s \to t} = R_{s \to t} C_t, 
\end{equation}
where $R_{s \to t}$ represents the equivalent resistance from $s$ to $t$, and $C_t$ represents the capacitance at node $t$. For net delay, $R$ and $C$ here are both linear to wirelength, making the delay quadratic in length. Thus, we choose the square of the pin-to-pin Euclidean distance -- quadratic loss as the objective:
\begin{equation}
    Q(i, j) = (x_i-x_j)^2 + (y_i-y_j)^2,
    \label{eq:Q_ij}
\end{equation}
where $Q(i, j)$ represents the pin-to-pin loss of pin $i$ and $j$.

Fig.~\ref{fig:critical_path} demonstrates the effectiveness of our quadratic loss design, comparing it to HPWL loss and Euclidean distance loss using the \texttt{superblue16} case~\cite{kim2015iccad}. We first identify the most critical path using \texttt{report\_timing(1)} from the coarse placement before timing optimization, as shown in Fig.~\ref{fig:critical_path}(a). Fig.~\ref{fig:critical_path}(b) and (c) show the corresponding path optimized to convergence by HPWL loss and linear Euclidean distance loss, respectively. Both paths and slacks appear similar in these two figures due to the nearly linear relationship both loss functions have with distance. Such linear loss fails to differentiate effectively between longer and shorter wires, with gradients indicating direction but not magnitude. Consequently, many cells may cluster together, while some wire segments become excessively long, as shown in Fig.~\ref{fig:critical_path}(b) and (c). In contrast, Fig.~\ref{fig:critical_path}(d), which utilizes a quadratic distance loss, demonstrates improved path slack despite an increase in total wirelength. This improvement is attributed to the quadratic loss fostering a more uniform distribution of cells and maintaining more consistent wire segment lengths. It also implies that there are fewer excessively long wire segments, which typically necessitate the insertion of buffers that can escalate area, power, and thermal concerns, underscoring the utility in modern chip design~\cite{ajayi2019toward}. As this study focuses on academic cases which are not suited for post-CTS evaluations, optimizing area, power, and thermal is identified as a crucial future work.

\begin{table*}[!htbp]
\renewrobustcmd{\bfseries}{\fontseries{b}\selectfont}
\renewrobustcmd{\boldmath}{}
\newrobustcmd{\B}{\bfseries}
\setlength{\tabcolsep}{4.2pt}
\setlength{\abovecaptionskip}{0.cm}
\setlength{\belowcaptionskip}{0.cm}
\centering
\begin{threeparttable}
\caption{Comparing TNS ($\times 10^5$ \si{ps}), WNS ($\times 10^3$ \si{ps}), and HPWL ($\times 10^6$) across different state-of-the-art timing-driven placement methods. The best results are in \textbf{bold}, and the runner-ups are colored \BR{brown}.}
\begin{tabular}{@{}l|ccc|ccc|ccc|ccc|ccc@{}}
\toprule
\multirow{2}{*}{\centering\arraybackslash Benchmark}
& \multicolumn{3}{c|}{DREAMPlace*~\cite{lin2019dreamplace}}
& \multicolumn{3}{c|}{DREAMPlace 4.0*~\cite{liao2023dreamplace}}
& \multicolumn{3}{c|}{Differentiable-TDP\textdagger~\cite{guo2022differentiable}}
& \multicolumn{3}{c|}{Distribution-TDP\S~\cite{lin2024timing}}         
& \multicolumn{3}{c}{Efficient-TDP (ours)} \\
& {TNS} & {WNS} & {HPWL} & {TNS} & {WNS} & {HPWL} & {TNS} & {WNS} & {HPWL} & {TNS} & {WNS} & {HPWL} & {TNS} & {WNS} & {HPWL} \\
\midrule
\texttt{superblue1}   & -262.44 & -18.87 & \BR{422.0}   & -85.03 & -14.10 & 443.1   & -74.85 & -10.77 & 432.8   & \BR{-42.10} & \BR{-9.26}  & - & \B -17.44 & \B -7.75 & \B 418.8 \\
\texttt{superblue3}   & -76.64  & -27.65 & \BR{478.2}   & -54.74 & -16.43 & 482.4   & -39.43 & -12.37 & 478.4   & \BR{-26.59} & \BR{-12.19} & - & \B -20.40 & \B -11.82 & \B 462.5 \\
\texttt{superblue4}   & -290.88 & -22.04 & \B 312.0   & -144.38& -12.78 & 335.9   & \BR{-82.92}  & \B -8.49  & \BR{312.2}   & -123.28& \BR{-8.86}  & - & \B -82.88 & -9.17 & 317.7 \\
\texttt{superblue5}   & -157.82 & -48.92 & \BR{488.3}   & -95.78 & -26.76 & 556.2   & -108.08 & \BR{-25.21} & 488.7   & \BR{-70.35} & -31.64 & - & \B -62.18 & \B -24.65 & \B 484.2 \\
\texttt{superblue7}   & -141.55 & -19.75 & 604.3   & -63.86 & \B -15.22 & 604.0   & \BR{-46.43}  & \B -15.22 & \BR{602.1}   & -95.89 & -17.24 & - & \B -43.52 & \B -15.22 & \B 597.5 \\
\texttt{superblue10}  & -731.94 & -26.10 & 935.9   & -768.75& -31.88 & 1036.7  & \B -558.05 & \B -21.97 & \BR{934.4}   & -691.10 & -25.86 & - & \BR{-558.14} & \BR{-23.08} & \B 911.6 \\
\texttt{superblue16}  & -453.57 & -17.71 & \B 435.8   & -124.18& -12.11 & \BR{448.1}   & -87.03  & \BR{-10.85} & 485.1   & \BR{-55.99}  & -12.21 & - & \B -22.90 & \B -8.63 & 471.6 \\
\texttt{superblue18}  & -96.76  & -20.29 & 243.0   & -47.25 & -11.87 & 253.6   & -19.31  & -7.99  & \BR{243.6}   & \BR{-19.23}  & \B -5.25  & - & \B -16.16 & \BR{-6.92} & \B 234.4 \\
\midrule
Average Ratio            & 6.90    & 2.07   & \BR{1.004} & 2.75   & 1.40   & 1.06 & 2.00    & \BR{1.09}   & 1.02 & \BR{1.68}    & 1.11   & - & \B 1.00 & \B 1.00 & \B 1.00 \\
\bottomrule
\end{tabular}
\label{table:main}
\begin{tablenotes}
\small
\item * For DREAMPlace~\cite{lin2019dreamplace} and DREAMPlace 4.0~\cite{liao2023dreamplace}, we replicate the layout DEFs using the default configurations.
\item \textdagger { }For Differentiable-TDP~\cite{guo2022differentiable}, we acquire the DEFs from its authors to evaluate.
\item \S { }For Distribution-TDP~\cite{lin2024timing}, we borrow their results as our evaluation script is identical. The HPWL values are not provided.
\end{tablenotes}
\end{threeparttable}
\end{table*}


\begin{table*}[t]
\renewrobustcmd{\bfseries}{\fontseries{b}\selectfont}
\renewrobustcmd{\boldmath}{}
\newrobustcmd{\B}{\bfseries}
\setlength{\tabcolsep}{7pt}
\centering
\caption{Ablation study comparing TNS ($\times 10^5$ \si{ps}) and WNS ($\times 10^3$ \si{ps}) across various settings. The best results are in \textbf{bold}, and the runner-ups are colored \BR{brown}.}
\begin{tabular}{@{}l|cc|cc|cc|cc|cc|cc@{}}
\toprule
\multirow{2}{*}{\centering\arraybackslash Benchmark}
& \multicolumn{2}{c|}{w/ HPWL Loss}
& \multicolumn{2}{c|}{w/ Linear Loss}
& \multicolumn{2}{c|}{w/ rpt\_timing(n*10)}
& \multicolumn{2}{c|}{w/ rpt\_timing\_ept(n,10)}
& \multicolumn{2}{c|}{w/o Path Extraction}
& \multicolumn{2}{c}{Our Method} \\
& {TNS} & {WNS} & {TNS} & {WNS} & {TNS} & {WNS} & {TNS} & {WNS} & {TNS} & {WNS} & {TNS} & {WNS} \\
\midrule
\texttt{superblue1}  & -74.70 & -13.85 & -76.66 & -11.94 & -80.61 & -9.84 & \B -12.69 & \BR{-8.79} & -19.27 & -14.43 & \BR{-17.44} & \B -7.75 \\
\texttt{superblue3}  & -47.42 & -15.81 & -47.29 & -13.00 & -37.10 & \B -11.71 & -20.93 & -11.91 & \B -20.11 & -16.47 & \BR{-20.40} & \BR{-11.82} \\
\texttt{superblue4}  & -155.23 & -16.35 & -153.76 & -14.32 & -139.05 & \BR{-9.15} & \BR{-86.49} & \B -8.75 & -102.39 & -10.23 & \B -82.88 & -9.17 \\
\texttt{superblue5}  & -93.21 & -26.37 & -91.96 & -28.23 & -123.13 & -27.28 & \BR{-58.78} & \BR{-25.58} & \B -51.61 & -34.57 & -62.18 & \B -24.65 \\
\texttt{superblue7}  & -68.68 & -16.19 & -59.47 & \B -15.22 & -47.70 & \B -15.22 & \BR{-35.14} & -19.57 & \B -34.96 & \B -15.22 & -43.52 & \B -15.22 \\
\texttt{superblue10} & -657.95 & -23.39 & -707.27 & -27.67 & -629.28 & -24.50 & -570.37 & -23.23 & \B -515.80 & \B -21.94 & \BR{-558.14} & \BR{-23.08} \\
\texttt{superblue16} & -61.96 & -9.93  & -63.69 & -13.81 & -30.87 & \BR{-9.11}  & -25.17 & -12.57 & \BR{-24.44} & -9.90 & \B -22.90 & \B -8.63 \\
\texttt{superblue18} & -51.62 & -13.18 & -48.21 & -13.70 & -34.69 & -7.40  & \B -15.19 & \BR{-7.20} & \BR{-15.38} & -7.64 & -16.16 & \B -6.92 \\
\midrule
Average Ratio           & 2.33 & 1.39 & 2.31 & 1.39 & 1.97 & \BR{1.07} & \B 0.95 & 1.12 & \BR{0.99} & 1.25 & 1.00 & \B 1.00 \\
\bottomrule
\end{tabular}
\label{table:ablation}
\end{table*}

\subsection{Workflow Summary}

We detail the overall process for timing optimization as a summary. Initially, vanilla DREAMPlace~\cite{lin2019dreamplace} is run to distribute the cells within the layout. Subsequently, we perform a path-level timing analysis every $m$ rounds to extract critical paths and update the pin-to-pin loss. This involves \texttt{report\_timing\_endpoint(n,1)}, where $n$ denotes the number of all failing endpoints, to collect data on critical paths. As we traverse these paths, each pin pair $(i, j)$ involved is added to a maintained set $P$, unless it has already been included. To address the path-sharing effect, the weight $w_{(i,j)}$ of each pin pair is dynamically updated as follows:
\begin{equation}
w_{(i,j)} =
\begin{cases}
w_0, & \text{if $(i, j)$ $\notin P$}, \\
w_{(i, j)} + w_1 \cdot (slack/{\text{WNS}}), & \text{otherwise},
\end{cases}
\label{eq:w_ij}
\end{equation}
where $w_0$ and $w_1$ are hyperparameters, and $slack$ indicates the negative slack of the respective critical path. The pin-to-pin attraction loss $PP(\bm{x}, \bm{y})$ of the layout is then computed as:
\begin{equation}
PP(\bm{x}, \bm{y}) = \sum_{(i,j) \in P} w_{(i,j)} \cdot Q(i,j),
\end{equation}
with $Q(i,j)$ and $w_{(i,j)}$ defined in Eqs.~\ref{eq:Q_ij} and~\ref{eq:w_ij}, respectively. After defining the loss function properly, we implement the CUDA kernel of $PP$ loss for GPU-acceleration.

\section{Experimental Results}\label{sec:experiments}

We have developed our timing-driven global placer based on the open-source placer DREAMPlace 4.0 released version\footnote{https://github.com/limbo018/DREAMPlace/releases/tag/4.0.0}. We assess the efficacy of our placer using the well-established benchmark suite from the ICCAD 2015 contest~\cite{kim2015iccad}. All the evaluations are conducted on a Linux server equipped with a 52-core Intel Xeon CPU at 2.60 GHz, an NVIDIA RTX 2080S GPU, and 128GB of RAM. The hyperparameters are set as follows: $\beta = 2.5 \times 10^{-5}$, $m=15$, $w_0 = 10$, and $w_1 = 0.2$. The updating rule for the Lagrange multiplier $\lambda$ is adopted from DREAMPlace. Timing optimization commences at the 500th iteration, a stage where cell distribution has typically stabilized, in line with the configurations specified in DREAMPlace 4.0.

\subsection{Main Results}

Table~\ref{table:main} presents a comprehensive comparison of TNS, WNS, and HPWL metrics between our timing-driven placer and four baseline methods. All DEF results are assessed using the official evaluation kit from the ICCAD 2015 contest to ensure fair comparison. Our approach significantly outperforms the state-of-the-art timing-driven placers, notably Differentiable-TDP~\cite{guo2022differentiable} and Distribution-TDP~\cite{lin2024timing}. Specifically, our method achieves the best TNS results in seven out of eight test cases, showing an average improvement of 50.0\% over Differentiable-TDP and 40.5\% over Distribution-TDP. Our placer also shows a consistent 8.3\% improvement in WNS compared to these two leading placers. Furthermore, when compared to DREAMPlace~\cite{lin2019dreamplace}, including its version 4.0~\cite{liao2023dreamplace}, our results consistently outperform theirs in all eight cases for TNS and WNS. It is surprising that our method surpasses baselines, including the original DREAMPlace, in terms of HPWL in six out of eight cases. This improvement can be attributed to our targeted pin-to-pin attraction strategy, which minimizes the impact on non-critical pins and effectively preserves wirelength quality, unlike DREAMPlace 4.0 which applies weights to numerous nets. What's more, additional timing-driven optimization iterations may further optimize HPWL against density, compared to DREAMPlace with an earlier convergence.

\subsection{Ablation Study}\label{sec:ablation}

Table~\ref{table:ablation} summarizes the ablation study. The first two columns, ‘w/ HPWL Loss’ and ‘w/ Linear Loss,’ replace the quadratic distance loss with HPWL and linear Euclidean losses, respectively. They both fall short of the quadratic loss, consistent with the discussion in Sec.~\ref{sec:quadratic_loss}. Nevertheless, they deliver a 15\% improvement in TNS over DREAMPlace 4.0~\cite{liao2023dreamplace}, demonstrating the effectiveness of our pin-to-pin attraction modeling and critical path extraction. Furthermore, the superior performance of quadratic loss compared to HPWL/Euclidean loss suggests an advantage over Electrostatics-TDP~\cite{lin2024electrostatics}, which relies on HPWL/Euclidean loss for its virtual path modeling. However, the differences in frameworks and datasets make direct comparisons with \cite{lin2024electrostatics} impossible.

The third column, ‘w/ rpt\_timing(n*10),’ uses OpenTimer’s original \texttt{report\_timing(n*10)} function for critical path extraction, where $n$ is the number of failing endpoints. As shown in Table~\ref{table:timing_statistics}, this approach provides insufficient coverage for comprehensive timing analysis, resulting in worse TNS performance. It also requires approximately $10\times$ more computation time compared to our path extraction method which uses \texttt{report\_timing\_endpoint(n,1)}.

The fourth column, ‘w/ rpt\_timing\_ept(n,10),’ extracts 10 critical paths (instead of one) per failing endpoint with \texttt{report\_timing\_endpoint(n,10)}. This adjustment improves TNS by incorporating more detailed timing information but slightly degrades WNS and increases computation time.

The fifth column, ‘w/o Path Extraction,’ replaces our path-level timing analysis with the pin-level timing information and momentum-based weighting scheme proposed by DREAMPlace~4.0~\cite{liao2023dreamplace}. While this method achieves competitive TNS results, it performs worse in WNS, likely because pin-level analysis fails to consider path-sharing effects, overlooking some critical paths that significantly influence timing.

\begin{table}[t]
\renewrobustcmd{\bfseries}{\fontseries{b}\selectfont}
\renewrobustcmd{\boldmath}{}
\setlength{\tabcolsep}{6pt}
\newrobustcmd{\B}{\bfseries}
\centering
\caption{Comparison on Runtime (\si{sec}) with DREAMPlace~\cite{lin2019dreamplace} and DREAMPlace 4.0~\cite{liao2023dreamplace}. The best results are in \textbf{bold}, and the runner-ups are colored \BR{brown}.}
\small 
\begin{tabular}{@{}l|c|c|c@{}}
\toprule
{\centering\arraybackslash Benchmark}
& {DREAMPlace} & {DREAMPlace 4.0} & {Our Method} \\
\midrule
\texttt{superblue1} & \B 122.95 & 615.61 & \BR{531.28} \\
\texttt{superblue3} & \B 125.34 & 798.10 & \BR{699.86} \\
\texttt{superblue4} & \B 80.88 & \BR{372.28} & 591.33 \\
\texttt{superblue5} & \B 147.35 & \BR{660.42} & 714.27 \\
\texttt{superblue7} & \B 190.32 & 926.91 & \BR{799.40} \\
\texttt{superblue10} & \B 252.49 & 1163.32 & \BR{1113.07} \\
\texttt{superblue16} & \B 62.61 & 442.99 & \BR{409.06} \\
\texttt{superblue18} & \B 56.89 & 368.06 & \BR{301.42} \\
\midrule
Average Ratio & \B 0.20 & 1.04 & \BR{1.00} \\
\bottomrule
\end{tabular}
\label{table:runtime}
\end{table}

\begin{figure}[t]
\centering
\includegraphics[width=0.42\textwidth]{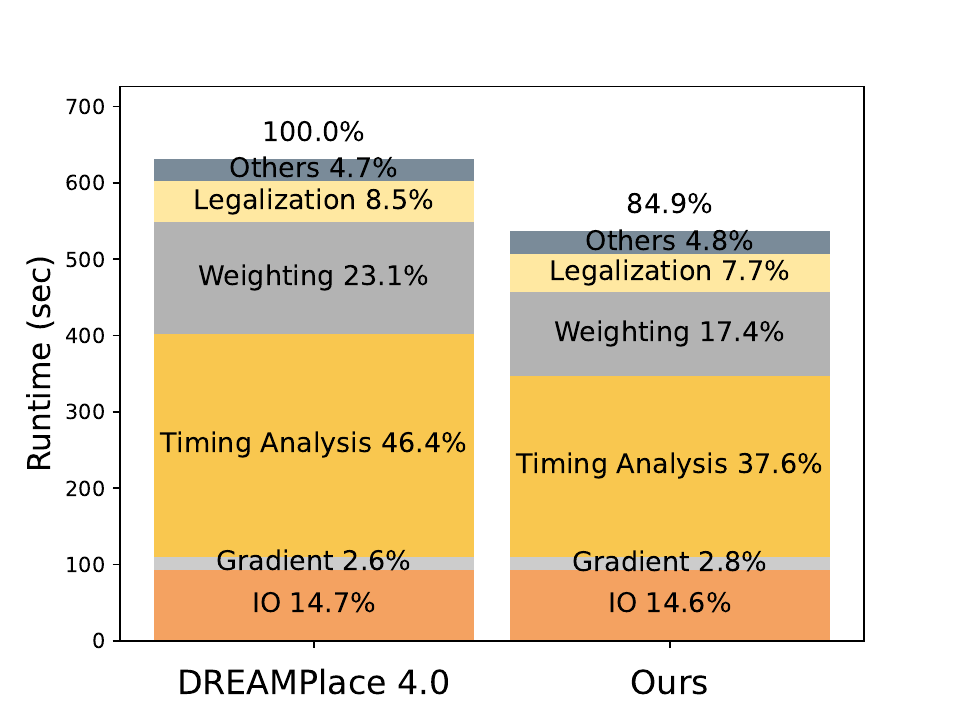}
\caption{Runtime breakdown comparison between DREAMPlace 4.0 and our method for case \texttt{superblue1}. The time spent by each component is normalized by 615 seconds, the total runtime of DREAMPlace 4.0.}
\label{fig:running_time_breakdown}
\end{figure}

\begin{figure}[t]
\centering
\includegraphics[width=0.45\textwidth]{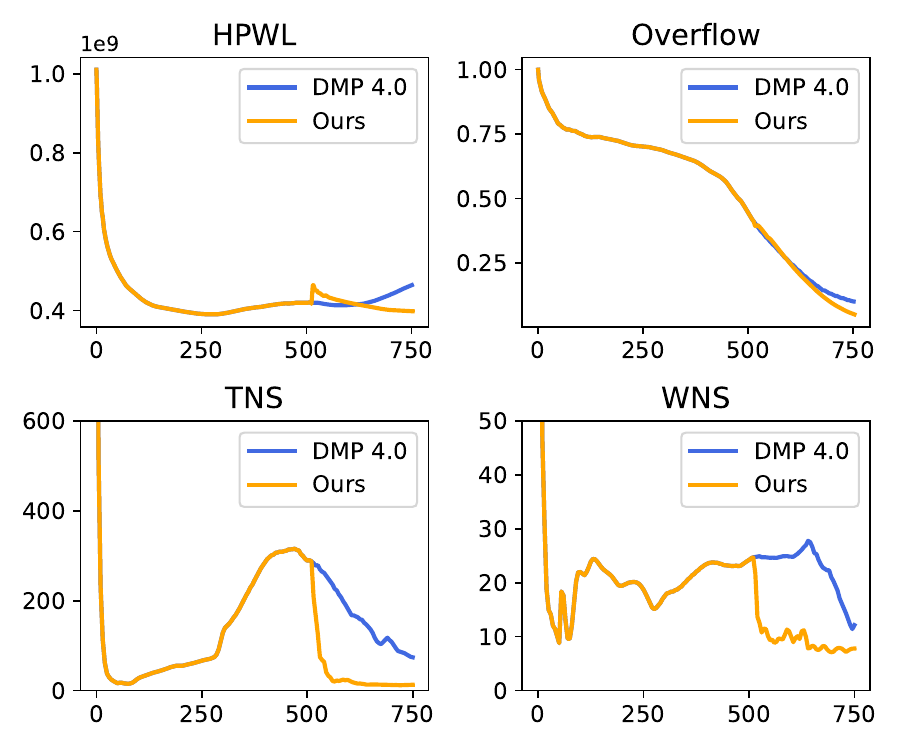}
\caption{Optimization iterations for case \texttt{superblue1}. The \textcolor{blue}{blue} curve is DREAMPlace 4.0, and the \textcolor{orange}{yellow} one is our method. Timing optimization of both methods starts from the \textbf{500th} iteration. TNS and WNS values are converted to their absolute values in the figure for better illustration.}
\label{fig:opt_curve}
\end{figure}

\subsection{Runtime Analysis and Additional Results}

Table~\ref{table:runtime} compares runtime among DREAMPlace~\cite{lin2019dreamplace}, DREAMPlace 4.0~\cite{liao2023dreamplace}, and our method across eight designs. DREAMPlace achieves the best runtime in all cases, as it focuses on wirelength without a timing engine which is time consuming. Our method surpasses DREAMPlace~4.0 in most cases thanks to our efficient timing analysis and weighting scheme, as illustrated in Fig.~\ref{fig:running_time_breakdown}. For case \texttt{superblue1}, we break down the runtime into key components and normalize each against DREAMPlace 4.0's total runtime for clarity. The reductions in our runtime primarily result from our efficient critical path extraction and pin pair weighting techniques.

Fig.~\ref{fig:opt_curve} depicts the HPWL, overflow, TNS, and WNS throughout a placement run, comparing our method with DREAMPlace 4.0~\cite{liao2023dreamplace}. The two curves align until the 500th iteration, at which point timing optimization commences. In the HPWL and Overflow sub-figures, the application of substantial net weights by DREAMPlace 4.0 leads to poorer HPWL performance and a slower convergence rate. Furthermore, our method rapidly enhances TNS and WNS performance, and maintains stability until the optimization fully converges, thereby demonstrating the efficacy of our timing objective design.

\section{Conclusion}\label{sec:conclusion}

In this work, we introduce a GPU-accelerated, timing-driven global placement framework that integrates a pin-to-pin attraction objective within the popular open-source DREAMPlace framework. We develop a novel critical path extraction method for rapid, precise timing analysis and design a quadratic distance loss function to closely align with the specific timing metrics, thereby enhancing our framework's performance. Experimentation on the ICCAD2015 benchmark suite shows substantial improvements over leading timing-driven placers.

\section{Acknowledgement}\label{sec::acknowledgement}

This work was supported by the National Science and Technology Major Project (2022ZD0116600), the National Science Foundation of China (62276124), and the Fundamental Research Funds for the Central Universities (14380020). The authors want to thank Zizheng Guo, Hao Kong and Liying Yang for their help and valuable discussions. Chao Qian is the corresponding author.


\clearpage
\clearpage
\bibliographystyle{IEEEtran}
\bibliography{main}

\end{document}